# Testing of sequences by simulation

Antal Iványi
Eötvös Loránd University
Department of Computer Algebra
H-1117, Budapest, Hungary
Pázmány sétány 1/C
email: tony@compalg.inf.elte.hu

Balázs Novák
Eötvös Loránd University
Department of Computer Algebra
H-1117, Budapest, Hungary
Pázmány sétány 1/C
email: psziho@inf.elte.hu

**Abstract.** Let $\xi$ be a random integer vector, having uniform distribution

$$\mathbf{P}\{\xi = (i_1, i_2, \ldots, i_n) = 1/n^n\} \text{ for } 1 \leq i_1, i_2, \ldots, i_n \leq n.$$

A realization $(i_1, i_2, \ldots, i_n)$ of $\xi$ is called *good*, if its elements are different. We present algorithms LINEAR, BACKWARD, FORWARD, TREE, GARBAGE, BUCKET which decide whether a given realization is good. We analyse the number of comparisons and running time of these algorithms using simulation gathering data on all possible inputs for small values of $n$ and generating random inputs for large values of $n$.

## 1 Introduction

Let $\xi$ be a random integer vector, having uniform distribution

$$P\{\xi = (i_1, i_2, \ldots, i_n)\} = 1/n^n$$

for $1 \leq i_1, i_2, \ldots, i_n \leq n$.

A realization $(i_1, i_2, \ldots, i_n)$ of $\xi$ is called *good*, if its elements are different. We present six algorithms which decide whether a given realization is good.

This problem arises in connection with the design of agricultural [4, 5, 57, 72] and industrial [34] experiments, with the testing of Latin [1, 9, 22, 23, 27, 32,

53, 54, 63, 64] and sudoku [3, 4, 6, 12, 13, 14, 15, 16, 17, 20, 21, 22, 26, 29, 30, 31, 41, 42, 44, 46, 47, 51, 55, 59, 61, 64, 66, 67, 68, 69, 70, 72, 74] squares, with genetic sequences and arrays [2, 7, 8, 18, 24, 28, 35, 36, 37, 38, 45, 48, 49, 50, 56, 65, 71, 73, 75], with sociology [25], and also with the analysis of the performance of computers with interleaved memory [11, 33, 39, 40, 41, 43, 52].

Section 2 contains the pseudocodes of the investigated algorithms. In Section 3 the results of the simulation experiments and the basic theoretical results are presented. Section 4 contains the summary of the paper.

Further simulation results are contained in [62]. The proofs of the lemmas and theorems can be found in [43].

## 2 Pseudocodes of the algorithms

This section contains the pseudocodes of the investigated algorithms LINEAR, BACKWARD, FORWARD, TREE, GARBAGE, and BUCKET. The psudocode conventions described in the book [19] written by Cormen, Leiserson, Rivest, and Stein are used.

The inputs of the following six algorithms are $n$ (the length of the sequence $s$) and $s = (s_1, s_2, \ldots, s_n)$, a sequence of nonnegative integers with $0 \leq s_i \leq n$ for $1 \leq i \leq n$) in all cases. The output is always a logical variable $g$ (its value is TRUE, if the input sequence is good, and FALSE otherwise).

The working variables are usually the cycle variables $i$ and $j$.

### 2.1 Definition of algorithm LINEAR

LINEAR writes zero into the elements of an $n$ length vector $\mathbf{v} = (v_1, v_2, \ldots, v_n)$, then investigates the elements of the realization and if $v[s_i] > 0$ (signalising a repetition), then stops, otherwise adds 1 to $v[s[i]]$.

LINEAR($n, s$)

```
01 g ← TRUE
02 for i ← 1 to n
03     do v[i] ← 0
04 for i ← 1 to n
05     do if v[s[i]] > 0
06         then g ← FALSE
07             return g
08         else  v[s[i]] ← v[s[i]] + 1
09 return g
```

## 2.2 Definition of algorithm BACKWARD

BACKWARD compares the second ($i_2$), third ($i_3$), ..., last ($i_n$) element of the realization **s** with the previous elements until the first collision or until the last pair of elements.

BACKWARD(n, **s**)

**01** $g \leftarrow$ TRUE
**02 for** $i \leftarrow 2$ **to** $n$
**03**     **do for** $j \leftarrow i - 1$ **downto** 1
**04**         **do if** $s[i] = s[j]$
**05**             **then** $g \leftarrow$ FALSE
**06**                 **return** $g$
**07 return** $g$

## 2.3 Definition of algorithm FORWARD

FORWARD compares the first ($s_1$), second ($s_2$), ..., last but one ($s_{n-1}$) element of the realization with the following elements until the first collision or until the last pair of elements.

FORWARD(n, **s**)

**01** $g \leftarrow$ TRUE
**02 for** $i \leftarrow 1$ **to** $n - 1$
**03**     **do for** $j \leftarrow i + 1$ **to** $n$
**04**         **do if** $s[i] = s[j]$
**05**             **then** $g \leftarrow$ FALSE
**06**                 **return** $g$
**07 return** $g$

## 2.4 Definition of algorithm TREE

TREE builds a random search tree from the elements of the realization and finishes the construction of the tree if it finds the following element of the realization in the tree (then the realization is not good) or it tested the last element too without a collision (then the realization is good).

TREE(n, **s**)

**01** $g \leftarrow$ TRUE
**02 let** $s[1]$ be the root of a tree
**03 for** $i \leftarrow 2$ **to** $n$

```
04      if [s[i] is in the tree
05         then g ← False
06               return
07         else  insert s[i] in the tree
08 return g
```

## 2.5 Definition of algorithm GARBAGE

This algorithm is similar to LINEAR, but it works without the setting zeros into the elements of a vector requiring linear amount of time.

Beside the cycle variable $i$ GARBAGE uses as working variable also a vector $\mathbf{v} = (v_1, v_2, \ldots, v_n)$. Interesting is that $\mathbf{v}$ is used without initialisation, that is its initial values can be arbitrary integer numbers.

The algorithm GARBAGE was proposed by Gábor Monostori [58].

GARBAGE$(n, \mathbf{s})$

```
01 g ← True
02 for i ← 1 to n
03     do if v[s[i]] < i and s[v[s[i]]] = s[i]
04            then g ← False
05                  return g
06            else  v[s[i]] ← i
07 return g
```

## 2.6 Definition of algorithm BUCKET

BUCKET handles the array $Q[1 : m, 1 : m]$ (where $m = \lceil \sqrt{n} \rceil$ and puts the element $s_i$ into the $r$th row of Q, where $r = \lceil s_i/m \rceil$ and it tests using linear search whether $s_j$ appeared earlier in the corresponding row. The elements of the vector $\mathbf{c} = (c_1, c_2, \ldots, c_m)$ are counters, where $c_j$ $(1 \leq j \leq m)$ shows the number of elements of the $i$th row.

For the simplicity we suppose that $n$ is a square.

BUCKET$(n, \mathbf{s})$

```
01 g ← True
02 m ← √n
03 for j ← 1 to m
04     do c[j] ← 1
05 for i ← 1 to n
06     do r ← ⌈s[i]/m⌉m
```

```
07          for j ← 1 to c[r] − 1
08              do if s[i] = Q[r, j]
09                  then g ← False
10                      return g
11                  else  Q[r, c[r]] ← s[i]
12                        c[r] ← c[r] + 1
13 return g
```

## 3 Analysis of the algorithms

### 3.1 Analysis of algorithm LINEAR

The first algorithm is LINEAR. It writes zero into the elements of an $n$ length vector $\mathbf{v} = (v_1, v_2, \ldots, v_n)$, then investigates the elements of the realization sequentially and if $i_j = k$, then adds 1 to $v_k$ and tests whether $v_k > 0$ signaling a repetition.

In best case LINEAR executes only two comparisons, but the initialization of the vector $\mathbf{v}$ requires $\Theta(n)$ assignments. It is called LINEAR, since its running time is $\Theta(n)$ in best, worst and so also in expected case.

**Theorem 1** *The expected number $C_{exp}(n, \text{LINEAR}) = C_L$ of comparisons of LINEAR is*

$$C_L = 1 - \frac{n!}{n^n} + \sum_{k=1}^{n} \frac{n!k^2}{(n-k)!n^{k+1}}$$

$$= \sqrt{\frac{\pi n}{2}} + \frac{2}{3} + \kappa(n) - \frac{n!}{n^n},$$

*where*

$$\kappa(n) = \frac{1}{3} - \sqrt{\frac{\pi n}{2}} + \sum_{k=1}^{n} \frac{n!k}{(n-k)!n^{k+1}}$$

*tends monotonically decreasing to zero when $n$ tends to infinity. $n!/n^n$ also tends monotonically decreasing to zero, but their difference $\delta(n) = \kappa(n) - n!/n^n$ is increasing for $1 \leq n \leq 8$ and is decreasing for $n \geq 8$.*

**Theorem 2** *The expected running time $T_{exp}(n, \text{LINEAR}) = T_L$ of LINEAR is*

$$T_L = n + \sqrt{2\pi n} + \frac{7}{3} + 2\delta(n),$$

| n | $C_L$ | $\sqrt{\pi n/2} + 2/3$ | $n!/n^n$ | $\kappa(n)$ | $\delta(n)$ |
|---|---|---|---|---|---|
| 1 | 1.000000 | 1.919981 | 1.000000 | 0.080019 | −0.919981 |
| 2 | 2.000000 | 2.439121 | 0.500000 | 0.060879 | −0.439121 |
| 3 | 2.666667 | 2.837470 | 0.222222 | 0.051418 | −0.170804 |
| 4 | 3.125000 | 3.173295 | 0.093750 | 0.045455 | −0.048295 |
| 5 | 3.472000 | 3.469162 | 0.038400 | 0.041238 | +0.002838 |
| 6 | 3.759259 | 3.736647 | 0.015432 | 0.038045 | +0.022612 |
| 7 | 4.012019 | 3.982624 | 0.006120 | 0.035515 | +0.029395 |
| 8 | 4.242615 | 4.211574 | 0.002403 | 0.033444 | +0.031040 |
| 9 | 4.457379 | 4.426609 | 0.000937 | 0.031707 | +0.030770 |
| 10 | 4.659853 | 4.629994 | 0.000363 | 0.030222 | +0.029859 |

Table 1: Values of $C_L$, $\sqrt{\pi n/2} + 2/3$, $n!/n^n$, $\kappa(n)$, and $\delta(n) = \kappa(n) - n!/n^n$ for $n = 1, 2, \ldots, 10$

where
$$\delta(n) = \kappa(n) - \frac{n!}{n^n}$$
*tends to zero when* $n$ *tends to infinity, further*

$$\delta(n+1) > \delta(n) \text{ for } 1 \leq n \leq 7 \text{ and } \delta(n+1) < \delta(n) \text{ for } n \geq 8.$$

Table 1 shows some concrete values connected with algorithm LINEAR.

### 3.2 Analysis of algorithm BACKWARD

The second algorithm is BACKWARD. This algorithm is a naive comparison-based one. BACKWARD compares the second ($i_2$), third ($i_3$), ..., last ($i_n$) element of the realization with the previous elements until the first repetition or until the last pair of elements.

The running time of BACKWARD is constant in the best case, but it is quadratic in the worst case.

**Theorem 3** *The expected number* $C_{\exp}(n, \text{BACKWARD}) = C_B$ *of comparisons of the algorithm* BACKWARD *is*

$$C_B = n + \sqrt{\frac{\pi n}{8}} + \frac{2}{3} - \alpha(n),$$

*where* $\alpha(n) = \kappa(n)/2 + (n!/n^n)((n+1)/2)$ *monotonically decreasing tends to zero when* $n$ *tends to* $\infty$.

Table 2 shows some concrete values characterizing algorithm BACKWARD.

| n | $C_B$ | $n - \sqrt{\pi n/8} + 2/3$ | $(n!/n^n)((n+1)/2)$ | $\kappa(n)$ | $\alpha(n)$ |
|---|---|---|---|---|---|
| 1 | 0.000000 | 1.040010 | 1.000000 | 0.080019 | 1.040010 |
| 2 | 1.000000 | 1.780440 | 0.750000 | 0.060879 | 0.780440 |
| 3 | 2.111111 | 2.581265 | 0.444444 | 0.051418 | 0.470154 |
| 4 | 3.156250 | 3.413353 | 0.234375 | 0.045455 | 0.257103 |
| 5 | 4.129600 | 4.265419 | 0.115200 | 0.041238 | 0.135819 |
| 6 | 5.058642 | 5.131677 | 0.054012 | 0.038045 | 0,073035 |
| 7 | 5.966451 | 6.008688 | 0.024480 | 0.035515 | 0.042237 |
| 8 | 6.866676 | 6.894213 | 0.010815 | 0.033444 | 0.027536 |
| 9 | 7.766159 | 7.786695 | 0.004683 | 0.031707 | 0.020537 |
| 10 | 8.667896 | 8.685003 | 0.001996 | 0.030222 | 0.017107 |

Table 2: Values of $C_B$, $n-\sqrt{\pi n/8}+2/3$, $(n!/n^n)((n+1)/2)$, $\kappa(n)$, and $\alpha(n) = \kappa(n)/2 + (n!/n^n)((n+1)/2)$ for $n = 1, 2, \ldots, 10$

The next assertion gives the expected running time of algorithm BACKWARD.

**Theorem 4** *The expected running time* $T_{exp}(n, \text{BACKWARD}) = T_B$ *of the algorithm* BACKWARD *is*

$$T_B = n + \sqrt{\frac{\pi n}{8}} + \frac{4}{3} - \alpha(n),$$

*where* $\alpha(n) = \kappa(n)/2 + (n!/n^n)((n+1)/2)$ *monotonically decreasing tends to zero when* $n$ *tends to* $\infty$.

### 3.3 Analysis of algorithm FORWARD

FORWARD compares the first ($s_1$), second ($s_2$), ..., last but one ($s_{n-1}$) element of the realization with the next elements until the first collision or until the last pair of elements.

Taking into account the number of the necessary comparisons in line 04 of FORWARD, we get $C_{best}(n, \text{FORWARD}) = 1 = \Theta(1)$, and $C_{worst}(n, \text{FORWARD}) = B(n, 2) = \Theta(n^2)$.

The next assertion gives the expected running time.

**Theorem 5** *The expected running time* $T_{exp}(n, \text{FORWARD}) = T_F$ *of the algorithm* FORWARD *is*

$$T_F = n + \Theta(\sqrt{n}). \tag{1}$$

Although the basic characteristics of FORWARD and BACKWARD are identical, as Table 3 shows, there is a small difference in the expected behaviour.

| $n$ | number of sequences | number of good sequences | $C_F$ | $C_W$ |
|---|---|---|---|---|
| 2 | 4 | 2 | 1.000000 | 1.000000 |
| 3 | 27 | 6 | 2.111111 | 2.111111 |
| 4 | 256 | 24 | 3.203125 | 3.156250 |
| 5 | 3 125 | 120 | 4.264000 | 4.126960 |
| 6 | 46 656 | 720 | 5.342341 | 5.058642 |
| 7 | 823 543 | 5 040 | 6.326760 | 5.966451 |
| 8 | 16 777 216 | 40 320 | 7.342926 | 6.866676 |
| 9 | 387 420 489 | 362 880 | 8.354165 | 7.766159 |

Table 3: Values of $n$, the number of possible input sequences, number of good sequences, expected number of comparisons of FORWARD ($C_F$) and expected number of comparisons of BACKWARD ($C_W$) for $n = 2, 3, \ldots, 9$

## 3.4   Analysis of algorithm TREE

TREE builds a random search tree from the elements of the realization and finishes the construction of the tree if it finds the following element of the realization in the tree (then the realization is not good) or it tested the last element too without a collision (then the realization is good).

The worst case running time of TREE appears when the input contains different elements in increasing or decreasing order. Then the result is $\Theta(n^2)$. The best case is when the first two elements of **s** are equal, so $C_{best}(n, \text{TREE}) = 1 = \Theta(1)$.

Using the known fact that the expected height of a random search tree is $\Theta(\lg n)$ we can get that the order of the expected running time is $\sqrt{n} \log n$.

**Theorem 6** *The expected running time* $T_T$ *of* TREE *is*

$$T_T = \Theta(\sqrt{n} \lg n). \tag{2}$$

| n | number of good inputs | number of comparisons | number of assignments |
|---|---|---|---|
| 1 | 100 000.000000 | 0.000000 | 1.000000 |
| 2 | 49 946.000000 | 1.000000 | 1.499460 |
| 3 | 22 243.000000 | 2.038960 | 1.889900 |
| 4 | 9 396.000000 | 2.921710 | 2.219390 |
| 5 | 3 723.000000 | 3.682710 | 2.511409 |
| 6 | 1 569.000000 | 4.352690 | 2.773160 |
| 7 | 620.000000 | 4.985280 | 3.021820 |
| 8 | 251.000000 | 5.590900 | 3.252989 |
| 9 | 104 | 6.148550 | 3.459510 |
| 10 | 33 | 6.704350 | 3.663749 |
| 11 | 17 | 7.271570 | 3.860450 |
| 12 | 3 | 7.779950 | 4.039530 |
| 13 | 3 | 8.314370 | 4.214370 |
| 14 | 0 | 8.824660 | 4.384480 |
| 15 | 2 | 9.302720 | 4.537880 |
| 16 | 0 | 9.840690 | 4.716760 |
| 17 | 0 | 10.287560 | 4.853530 |
| 18 | 0 | 10.719770 | 4.989370 |
| 19 | 0 | 11.242740 | 5.147560 |
| 20 | 0 | 11.689660 | 5.279180 |

Table 4: Values of $n$, number of good inputs, number of comparisons, number of assignments of TREE for $n = 1, 2, \ldots, 10$

Table 4 shows some results of the simulation experiments (the number of random input sequences is 100 000 in all cases).

Using the method of the smallest squares to find the parameters of the formula $a\sqrt{n}\log_2 n$ we received the following approximation formula for the expected number of comparisons:

$$C_{exp}(n, \text{TREE}) = 1.245754\sqrt{n}\log_2 n - 0.273588.$$

### 3.5 Analysis of algorithm GARBAGE

This algorithm is similar to LINEAR, but it works without the setting zeros into the elements of a vector requiring linear amount of time.

Beside the cycle variable $i$ GARBAGE uses as working variable also a vector

$\mathbf{v} = (v_1, v_2, \ldots, v_n)$. Interesting is that $\mathbf{v}$ is used without initialisation, that is its initial values can be arbitrary integer numbers.

The worst case running time of GARBAGE appears when the input contains different elements and the garbage in the memory does not help, but even in this case $C_{worst}(n, \text{GARBAGE}) = \Theta(n)$. The best case is when the first element is repeated in the input and the garbage helps to find a repetition of the firs element of the input. Taking into account this case we get $C_{best}(n, \text{GARBAGE}) = \Theta(1)$.

According to the next assertion the expected running time is $\Theta(\sqrt{n})$.

**Lemma 7** *The expected running time of* GARBAGE *is*

$$T_{exp}(n, \text{GARBAGE}) = \Theta(\sqrt{n}). \tag{3}$$

### 3.6  Analysis of algorithm BUCKET

Algorithm BUCKET divides the interval $[1, n]$ into $m = \lceil \sqrt{n} \rceil$ subintervals $I_1, I_2, \ldots, I_m$, where $I_k = [(k-1)m + 1, km)]$, and assigns a bucket $B_k$ to interval $I_k$. BUCKET sequentially puts the input elements $i_j$ into the corresponding bucket: if $i_j$ belongs to the interval $I_k$ then it checks whether $i_j$ is contained in $B_k$ or not. BUCKET works up to the first repetition. (For the simplicity we suppose that $n = m^2$.)

In best case BUCKET executes only 1 comparison, but the initialization of the buckets requires $\Theta(\sqrt{n})$ assignments, therefore the best running time is also $\sqrt{n}$. The worst case appears when the input is a permutation. Then each bucket requires $\Theta(n)$ comparisons, so the worst running time is $\Theta(n\sqrt{n})$.

**Lemma 8** *Let* $b_j$ *(*$j = 1, 2, \ldots, m$*) be a random variable characterising the number of elements in the bucket* $B_j$ *at the moment of the first repetition. Then*

$$E\{b_j\} = \sqrt{\frac{\pi}{2}} - \mu(n)$$

*for* $j = 1, 2, \ldots, m$, *where*

$$\mu(n) = \frac{1}{3\sqrt{n}} - \frac{\kappa(n)}{\sqrt{n}},$$

*and* $\mu(n)$ *tends monotonically decreasing to zero when* $n$ *tends to infinity.*

Table 5 contains some concrete values connected with $E\{b_1\}$.

| n | E{b₁} | $\sqrt{\pi/2}$ | $1/(3\sqrt{n})$ | $\kappa(n)/\sqrt{n}$ | $\mu(n)$ |
|---|-------|----------------|-----------------|----------------------|----------|
| 1 | 1.000000 | 1.253314 | 0.333333 | 0.080019 | 0.253314 |
| 2 | 1.060660 | 1.253314 | 0.235702 | 0.043048 | 0.192654 |
| 3 | 1.090055 | 1.253314 | 0.192450 | 0.029686 | 0.162764 |
| 4 | 1.109375 | 1.253314 | 0.166667 | 0.022727 | 0.143940 |
| 5 | 1.122685 | 1.253314 | 0.149071 | 0.018442 | 0.130629 |
| 6 | 1.132763 | 1.253314 | 0.136083 | 0.015532 | 0.120551 |
| 7 | 1.147287 | 1.253314 | 0.125988 | 0.013423 | 0.112565 |
| 8 | 1.147287 | 1.253314 | 0.117851 | 0.011824 | 0.106027 |
| 9 | 1.152772 | 1.253314 | 0.111111 | 0.010569 | 0.100542 |
| 10 | 1.157462 | 1.253314 | 0.105409 | 0.009557 | 0.095852 |

Table 5: Values of E{$b_1$}, $\sqrt{\pi/2}$, $1/(3\sqrt{n})$, $\kappa(n)/\sqrt{n}$, and $\mu(n) = 1/(3\sqrt{n}) - \kappa(n)/\sqrt{n}$ of BUCKET for $n = 1, 2, \ldots, 10$

**Lemma 9** *Let $f_n$ be a random variable characterising the number of comparisons executed in connection with the first repeated element. Then*

$$E\{f_n\} = 1 + \sqrt{\frac{\pi}{8}} - \eta(n),$$

*where*

$$\eta(n) = \frac{\frac{1}{3} + \sqrt{\frac{\pi}{8}} - \frac{\kappa(n)}{2}}{\sqrt{n} + 2},$$

*and $\eta(n)$ tends monotonically decreasing to zero when $n$ tends to infinity.*

**Theorem 10** *The expected number $C_{exp}(n, \text{BUCKET}) = C_B$ of comparisons of algorithm* BUCKET *in 1 bucket is*

$$C_B = \sqrt{n} + \frac{1}{3} - \sqrt{\frac{\pi}{8}} + \rho(n),$$

*where*

$$\rho(n) = \frac{5/6 - \sqrt{9\pi/8} - 3\kappa(n)/2}{\sqrt{n} + 1}$$

*tends to zero when $n$ and tends to infinity.*

| Index and Algorithm | $C_{best}(n)$ | $C_{worst}(n)$ | $C_{exp}(n)$ |
|---|---|---|---|
| 1. LINEAR | $\Theta(1)$ | $\Theta(n)$ | $\Theta(\sqrt{n})$ |
| 2. BACKWARD | $\Theta(1)$ | $\Theta(n^2)$ | $\Theta(n)$ |
| 3. FORWARD | $\Theta(1)$ | $\Theta(n^2)$ | $\Theta(n)$ |
| 4. TREE | $\Theta(1)$ | $\Theta(n^2)$ | $\Theta(\sqrt{n}\lg n)$ |
| 5. GARBAGE | $\Theta(1)$ | $\Theta(n)$ | $\Theta(\sqrt{n})$ |
| 6. BUCKET | $\Theta(\sqrt{n})$ | $\Theta(n\sqrt{n})$ | $\Theta(\sqrt{n})$ |

Table 6: The number of necessary comparisons of the investigated algorithms in best, worst and expected cases

**Theorem 11** *The expected running time $T_B(n, \text{BUCKET}) = T_B$ of BUCKET is*

$$T_B = \left(3 + 3\sqrt{\frac{\pi}{2}}\right)\sqrt{n} + \sqrt{\frac{25\pi}{8}} + \varphi(n),$$

*where*

$$\varphi(n) = 3\kappa(n) - \rho(n) - 3\eta(n) - \frac{n!}{n^n} - \frac{3\sqrt{\pi/8} - 1/3 - 3\kappa(n)/2}{\sqrt{n} + 1}$$

*and $\varphi(n)$ tends to zero when $n$ tends to infinity.*

It is worth to remark that simulation experiments of B. Novák [62] show that the expected running time of GARBAGE is a few percent better, then the expected running time of BUCKET.

## 4 Summary

Table 6 contains the number of necessary comparisons in best, worst and expected cases for all investigated algorithms.
Table 7 contains the running time in best, worst and expected cases for all investigated algorithms.

**Acknowledgements.** The authors thank Tamás F. Móri [60] for proving Lemma 8 and 9 and Péter Burcsi [10] for useful information on references, both are teachers of Eötvös Loránd University.

The European Union and the European Social Fund have provided financial support to the project under the grant agreement no. TÁMOP 4.2.1/B-09/1/KMR-2010-0003.

| Index and Algorithm | $T_{best}(n)$ | $T_{worst}(n)$ | $T_{exp}(n)$ |
|---|---|---|---|
| 1. Linear | $\Theta(n)$ | $\Theta(n)$ | $n + \Theta(\sqrt{n})$ |
| 2. Backward | $\Theta(1)$ | $\Theta(n^2)$ | $\Theta(n)$ |
| 3. Forward | $\Theta(1)$ | $\Theta(n^2)$ | $\Theta(n)$ |
| 5. Tree | $\Theta(1)$ | $\Theta(n^2)$ | $\Theta(\sqrt{n}\lg n)$ |
| 6. Garbage | $\Theta(1)$ | $\Theta(n)$ | $\Theta(\sqrt{n})$ |
| 7. Bucket | $\Theta(\sqrt{n})$ | $\Theta(n\sqrt{n})$ | $\Theta(\sqrt{n})$ |

Table 7: The running times of the investigated algorithms in best, worst and expected cases